\documentclass[namedreferences]{SolarPhysics}
\usepackage[optionalrh]{spr-sola-addons} 
\usepackage{solaheader}
\newcommand{\solareceiveddate}{1 October 2009} 
\newcommand{\solaaccepteddate}{9 December 2009}
\usepackage{graphicx}                    
\usepackage{amssymb}                    
\usepackage{color}                       
\usepackage{url}                         
\usepackage{enumerate}

\begin{document}
\begin{article}
\begin{opening}
\title{Structural Invariance of Sunspot Umbrae Over the Solar Cycle: 1993\,--\,2004}
%
\author{T.A.~\surname{Schad}$^{1,2}$\sep
       M.J.~\surname{Penn}$^{1}$
       }
%
\runningauthor{T.A. Schad, M.J. Penn}
\runningtitle{Umbral Structure Over the Solar Cycle}
%
   \institute{$^{1}$ National Solar Observatory,
	    \thanks{Operated by the Association of Universities for
        	  Research in Astronomy, Inc. (AURA), under cooperative 
	          agreement with the National Science Foundation} 
		     950 N Cherry Ave, Tucson AZ 85719,USA
                     email: \url{schad@noao.edu} email: \url{mpenn@noao.edu} \\
	     $^{2}$Department of Planetary Sciences, University of Arizona, Tucson, AZ 85721,USA
           }
\begin{abstract}
Measurements of maximum magnetic flux, minimum intensity, and size are presented for 12\,967 sunspot umbrae detected on the NASA/NSO spectromagnetograms between 1993 and 2004 to study umbral structure and strength during the solar cycle.  The umbrae are selected using an automated thresholding technique.  Measured umbral intensities are first corrected for a confirming observation of umbral limb-darkening.  Log--normal fits to the observed size distribution confirm that the size spectrum shape does not vary with time.  The intensity--magnetic flux relationship is found to be steady over the solar cycle.  The dependence of umbral size on the magnetic flux and minimum intensity are also independent of cycle phase and give linear and quadratic relations, respectively.  While the large sample size does show a low amplitude oscillation in the mean minimum intensity and maximum magnetic flux correlated with the solar cycle, this can be explained in terms of variations in the mean umbral size.  These size variations, however, are small and do not substantiate a meaningful change in the size spectrum of the umbrae generated by the Sun.  Thus, in contrast to previous reports, the observations suggest the equilibrium structure, as testified by the invariant size-magnetic field relationship, as well as the mean size (\textit{i.e.} strength) of sunspot umbrae do not significantly depend on solar cycle phase.
\end{abstract}
\end{opening}
%
\section{Introduction}\label{s:intro}

Magnetohydrostatic equilibrium describes the balance of the magnetic and thermal pressures across individual sunspots \cite{alfven43}. Along the direction parallel to the solar surface, the local environment of a sunspot markedly shifts from the cooler, convection inhibited, umbral region, hosting multi-kilogauss magnetic fields, to the convectively turbulent, hotter, external plasma of the quiet sun with considerably lower magnetic flux density. The mechanism responsible for this inverse relationship between magnetic field strength and temperature remains unclear \cite{biermann41,chou87} since umbrae host a wide range of thermal-magnetic field distributions \cite{pilvaz93,leonard08}. Interesting, however, is the empirical relationship between peak magnetic field strength and minimum intensity for different umbrae \cite{deinzer65,kopp92,livingston02}. \inlinecite{nort04} support a single power law to express this relationship. As such, the total magnetic pressure of a sunspot always \textit{more than} compensates for the corresponding reduction in thermal pressure, leading to a predictably larger spatial extent for umbrae with greater field strengths \cite{brants82,kopp92} and lower minimum intensities \cite{mathew07,wes08}.  However, to date, no study of the temporal stability of the relationship claimed by Norton and Gilman has been completed for a statistically significant number of umbrae.

We are interested in whether this equilibrium structure is influenced by solar cycle progression.   The recent study of 3931 umbrae by \inlinecite{penn07} revealed a solar-cycle oscillation in average umbral intensity between 1992 and 2003.  This agreed well with the mean intensity variation observed by \inlinecite{nort04} as well as both the mean intensity and magnetic field changes reported by \inlinecite{penn06}; though, both of those reports neglected the necessary size parameter, as addressed by \inlinecite{mathew07}.  Importantly, the intensity oscillation reported by \citeauthor{penn07} occurs without any evident corresponding umbral size variation. This thus agrees with the constant umbral size spectrum shape observed over multiple solar cycles by \inlinecite{bogdan88}.  Since the observed relationship between intensity and size did not change, the reported mean intensity oscillation insinuates that the magnetic field-size relationship changes over the solar cycle if the relationship claimed by \citeauthor{nort04} is indeed constant.  A cycle variation in the umbral magnetic twist might influence this relationship, but has not yet been observed \cite{pevtsov01}.

As this is the only significant evidence for the local magnetic properties of sunspot umbrae being influenced by the evolution of the global solar field, further exploration is imperative as it may have implications to both the solar dynamo \cite{schussler80,nort04} and umbral magnetic field orientation \cite{ruedi95,pevtsov01}.  It is worth noting, though, that \inlinecite{livingston06} reported a significantly lower number of sunspot groups with strong fields in odd numbered solar cycles compared to each cycle's preceding even numbered cycle pointing to a possible solar cycle influence on sunspot group generation.  We address this topic more completely than previous authors by revisiting the data from the Kitt-Peak spectromagnetograms (Section~\ref{s:data}) used by \inlinecite{penn07} to exploit its measurements of magnetic field for the first time in tandem with the intensity measurements.  Automating our sunspot detection (Section~\ref{s:selection}) yields the greatest number of umbrae so far measured for this type of study.  We explore the temporal variability of the relationships between peak magnetic field strength, minimum intensity, and umbral size (Section~\ref{s:results}) to discuss the alleged dependence of sunspot structure on solar cycle progression.

\section{Instrument and Data Description}

\subsection{The NASA/NSO Spectromagnetograph}\label{s:data}

We analyze 2771 full-disk daily maps of line-of-sight (LOS) magnetic flux density and continuum intensity produced by the NASA/NSO Spectromagnetograph (SPM) \cite{jones92} between 1993 and 2004 at the Kitt Peak Vacuum Telescope (KPVT).  Each map results from the on-line analysis of scanned long-slit spectra in both states of circular polarization centered on the 8688 \AA\ Fe {\sc i} absorption line (Land\'e g $=$ 1.66) as described in \inlinecite{jones96}.  Values for LOS magnetic flux and continuum intensity are therefore spatially and temporally synchronous. The full-disk maps have a field of view of $34'\times34'$ and a pixel size of $1.14''\times1.14''$.  

LOS magnetic flux measurements from SPM are utilized despite problems alluded to by \inlinecite{penn07}.  New observations from the spectropolarimeter onboard Hinode \cite{moon07} as well as more precise infrared measurements \cite{kopp92,livingston02} have shown that many synoptic magnetographs have underestimated magnetic flux, especially within active regions.  MDI underestimates LOS field values in the penumbra by almost a factor of two \cite{leonard08}.  Values from SPM compare well with GONG+, but both are 20\,--\,40 percent lower than MDI \cite{thornton02}.  In accordance with \inlinecite{jones01}, we multiply the SPM magnetic flux values by 1.41 to calibrate their values with MDI.  We proceed with our use of the SPM magnetic flux values since they are crucial to the study at hand, but we are careful to compare their results with other measurements.

As always, stray light is a considerable problem when discussing umbral intensity and it also can affect the determination of the magnetic field.  We fully acknowledge that for a more precise description of the intensity of a sunspot, a stray light correction is necessary.  Daily stray-light corrections are unfortunately impossible with the SPM for it is a scanning slit device in which the limb is observed at different times than the center of the disk.  \inlinecite{penn07}, though, studied the long term variation of stray-light within the instrument and found a very weak temporal trend (about 0.5\%).
Furthermore, as the SPM measures the continuum at infrared wavelengths, intensity measurements are expected to be less affected by stray-light than instruments operating at visible wavelengths.  Thus, while stray light does affect our complete determination of the sunspot intensity, its effect is constant enough to allow us to comment on its solar cycle variability.

\subsection{Detection of Sunspot Umbra}\label{s:selection}

We identify umbrae on the full-disk SPM continuum maps using a thresholding method \cite{brant90} with a single definition for the umbra-penumbra boundary as in \inlinecite{mathew07} and \inlinecite{wes08}.  The contrast of features on each map is the ratio of the measured intensity to the local intensity.  We define the local intensity using the Legendre polynomial fits of observed continuum intensity with respect to the cosine of the heliocentric angle ($\mu$) which are recorded in each data file header.  This method reduces inconsistencies in determining the local photospheric intensity due to large multi-umbrae active regions as well as local brightenings due to diverted energy flux and plage regions around the sunspot \cite{rast99}.  Our umbra-penumbra boundary definition is set at a contrast value of 0.62 for the SPM maps of continuum intensity near 8688 \AA, which is slightly lower than that used by \inlinecite{mathew07}, 0.655, and \inlinecite{wes08}, 0.68.  This differs also from the variable umbra-penumbra boundary defined by \inlinecite{penn07}, which will be discussed in Section~\ref{s:conclusions}.  

Our method for including complex spots is consistent with the approach used by other authors \cite{livingston09}.  Multiple umbrae that share the same penumbra within single complex sunspots are treated as individual umbrae.  We require these umbrae to be separated by at least one $1.14''$ arcsec pixel with a penumbral intensity value, and each umbra must contain at least two observable pixels with umbral intensity values.  Our conservative intensity threshold value counts connected umbrae as one.  Furthermore, we find only 4.5\% of the umbrae selected to be closely spaced, where we consider 'closely spaced' to mean the smaller umbrae of a two-umbrae pair is less than two times its radius away from the nearest pixel of the larger umbra.  We, in general, do not attempt to locate umbrae that are close to the limb ($\mu<0.4$) since magnetic field values are less reliable, projection effects are difficult to remove consistently, and the Wilson depression conceals a portion of the umbrae.  We also verify the selection of each umbra using the synchronous maps of magnetic field. We require that each umbra have a peak magnetic flux greater than 50 gauss, which is more than three sigma deviations from the background noise of the SPM magnetograms.  In total, we locate 12967 umbrae between 1993 and 2003.

For each umbra we glean values for the umbral area and radius, as well as the minimum continuum intensity and its 
corresponding magnetic flux density.  The umbral area simply corresponds to its spatial coverage by counting the number of pixels observed and correcting for angular foreshortening.  We calculate an ''effective'' umbral radius (later referred to as simply the radius) by assuming that each umbra is circular.  Using the solar ephemeris for each individual SPM map, angular units are converted into physical units (10$^{3}$ km).  The use of microhemispheres in Section~\ref{s:size_dist} is also corrected for the annual variation in the apparent solar semidiameter.  All values of LOS magnetic flux and continuum intensity presented here reflect the value measured at the darkest point in each sunspot umbra, which nearly always corresponds to the peak umbral magnetic field strength.  This dark umbral core evolves the least during its formation and evolution (Socas-Navarro 2003) and is least affected by umbral dots, which affect the intensity distribution of the umbra but only minimally affect the determined umbral radius.  Furthermore, in the dark umbral core the magnetic field is known to be the most vertically directed which allows us to adopt the vertical field approximation,
\begin{equation}\label{eq:vert_b}
{B_{r}=\frac{B_{LOS}}{\mu}}
\end{equation}
where $\mu = cos (\theta)$.  The heliocentric angle, $\theta$, is the angular distance between the observer's line of sight and the local radial vector.

\section{Results and Discussion}\label{s:results}

In choosing to measure minimum intensity (\textbf{I}), corresponding magnetic flux density (\textbf{B}), and umbral size (\textbf{R}), we parameterize a much more complicated system into merely three interdependent variables.  We are interested not only in whether the interdependencies of these variables change over the solar cycle, but also whether the average umbral intensity and magnetic field vary as suggested by \inlinecite{penn07}. For this, we begin with a size spectrum analysis of the observed umbrae to address whether the solar cycle influences the distribution of umbrae generated at a particular phase. A regression analysis is then performed to test the temporal stability of the B--I, B--R, and I--R relationships.  We finish by showing the temporal variation of the mean intensity and magnetic field as in \citeauthor{penn07} and carefully use the relationships derived to show that they are consistent with a temporal variation in size.

As a much greater number of smaller umbrae appear on the solar disk than larger umbrae, our regression analysis requires binning to ensure that the smaller spots do not dominate the information available in larger spots.  Thus, prior to performing any regression analysis, the data is binned using the horizontal (independent) variable into bins of equal width. Using the Levenberg-Marquardt least-squares procedure, the regression is derived between the mean values of the independent and dependent variables. The standard error of each mean is also carefully propagated into the analysis and is reflected in the regression error.

\subsection{Size Spectrum Analysis}\label{s:size_dist}

\begin{figure} 
\centerline{\includegraphics[width=0.95\textwidth,clip=]{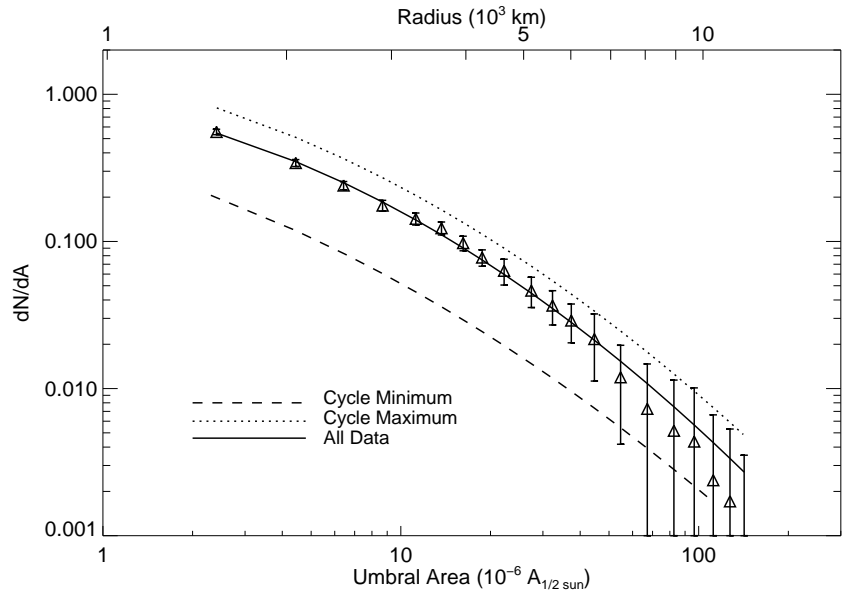}}
\caption{Comparison of the lognormal umbral size spectrum during cycle minimum, cycle maximum, and for all SPM data. The data points show the overall size spectrum.  The error bar magnitude is set by the square root of the number of spots in each bin. See text and Table~\ref{tbl:lognormal} for lognormal fit description and derived parameters.}\label{fig:size_dist}
\end{figure}
 
An analysis of the spectrum of umbral sizes has been carried out in the manner described by \inlinecite{bogdan88}.  It is important to address the spectrum of umbral sizes over the solar cycle rather than merely the average size. Deriving the size spectrum of umbrae has traditionally been used to comment on the generating mechanism for sunspot umbrae, whether that be flux-tube fragmentation or coalescence.  In a similar fashion, we utilize the size spectrum as a means to explore the influence of the solar cycle on the generation of umbrae.  As \citeauthor{bogdan88} remark, however, the spectrum derived is not identical to the distribution of umbral areas at a particular time.  Rather, it is the statistical probability of locating a spot of a particular size on the solar disk at any given time.  Thus, it cannot account for daily, monthly, or even annual variations of the mean umbral size, especially during solar minimum when solar activity is reduced.  Furthermore, unlike mere averaging of umbral sizes, the spectrum is influenced less by the ability to consistently select small spots (comparable in size to the atmospheric seeing limit). The report of \inlinecite{bogdan88} gave convincing evidence for a stable size spectrum, independent of cycle phase, between 1917-1982.  Though, since we are studying a solar cycle not included in that study, it is important to address the spectrum observed for we cannot assume one cycle is identical to another.  For our selected umbrae, we follow nearly the same procedure as \citeauthor{bogdan88} to determine the log-normal distribution defined by
\begin{equation}
{\mathrm{ln}\left(\frac{\mathrm{dN}}{\mathrm{dA}}\right)=-\frac{(\mathrm{\ln A}-\mathrm{\ln\langle A \rangle)^{2}}}{2\ln\sigma_{A}}+\ln\left ( \frac{\mathrm{dN}}{\mathrm{dA}} \right )_{max}}
\end{equation}
where A is the umbral area (in microhemispheres), $\langle A \rangle$ is the spectrum mean, $\sigma_{A}$ is the geometric standard deviation, and dN/dA is the size distribution function.  The distribution function and its error, which is conservatively taken to be the square root of the number of umbrae in each bin, is divided by the number of maps measured as in the \citeauthor{bogdan88} study.  We employ the Levenberg-Marquardt least-squares fit procedure to determine the fit parameters for our observed size distribution over different portions of the solar cycle.  The error in the observed distribution is propagated into the fit.  
\begin{table}
\caption{Size Spectrum Lognormal Fit Parameters}\label{tbl:lognormal}
\begin{tabular}{lcrrrrr}
\hline
Cycle Phase & N spots & $\langle A\rangle$ & $\langle A\rangle_{err}$ & $\sigma_{A}$ & $\sigma_{A,err}$ & $\left(\frac{d\mathrm{N}}{d\mathrm{A}}\right)_{max}$ \\ 
\hline
Falling (Mar 93 - Sep 96)  &  1123 & 0.5269 & 0.6833 & 12.2519 & 14.8800 & 0.3933 \\
Rising (Oct 96 - Mar 01)   &  5127 & 0.5765 & 0.3734 & 12.7540 &  8.2612 & 0.8630 \\
Falling (Apr 01 - Sep 03)  &  4423 & 0.2022 & 0.2701 & 36.5095 & 48.4679 & 1.6876 \\
Minimum (1995 - 1997)      &   543 & 0.1032 & 0.4765 & 59.3642 & 260.715 & 0.6699 \\
Maximum (2000 - 2002)      &  5739 & 0.2394 & 0.2517 & 31.7978 & 33.3728 & 1.7479 \\
All data (1993 - 2003)     & 10673 & 0.3577 & 0.2210 & 20.7520 & 12.7818 & 0.9923 \\
\hline
\end{tabular}
\end{table}
Unlike \citeauthor{bogdan88} however, we do not limit our spots to those observed within 7.5 degrees of the central meridian. This maintains a statistically relevant number of umbrae from the shorter time frame of the SPM data. Our fits then are not expected to follow those of \citeauthor{bogdan88} precisely; however, it does grant us a measure of the solar cycle variability of our selected umbrae. The observed spectrum of the 10673 umbrae with area greater than the 1.5 $\mu$hemispheres seeing and resolution limit imposed by \citeauthor{bogdan88}, is presented in Figure~\ref{fig:size_dist}.  Included in the figure are the calculated log-normal distribution fits for all umbrae and umbrae of solar minimum and solar maximum. The size spectrum is not observed to vary in shape significantly over the solar cycle and the geometric mean of each fit is within each fit's one sigma error, which is consistent with \citeauthor{bogdan88}.  While the geometric mean of the spectrum remains constant, however, we note in Section~\ref{s:cycle} that the average umbral radius (binned annually) varies from year to year within our data set. The total range of these points is approximately $\pm486$ km from the mean of the unbinned data.  The error in the geometric mean of the observed size spectrum (equivalent to a radius change of approximately 462 km) identifies all but one data point as less than a one standard deviation from the determined mean of the steady size spectrum.  This supports our determination of the umbral size spectrum but does not discount the significant variations in mean umbral radius that we observe, which will be discussed in Section~\ref{s:cycle}.

\subsection{Center-To-Limb Intensity Variation}

\begin{figure} 
\centerline{\includegraphics[width=0.95\textwidth,clip=]{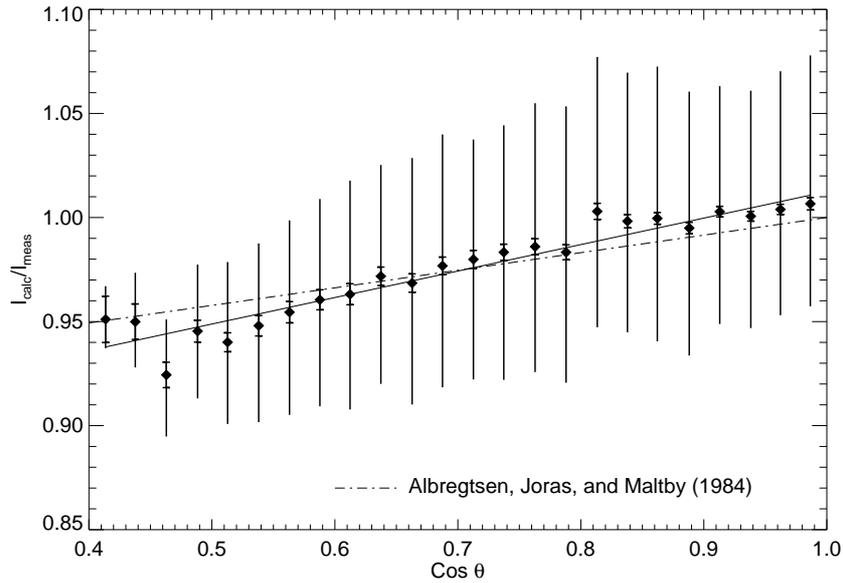}}
\caption{The center-to-limb dependence of umbral intensity shown as a ratio of the calculated intensity of each umbra inferred from its size to its true measured intensity.  Each data point is the mean value of points within a horizontal bin 0.025 wide. Smaller hatted error bars exhibit the standard error of the mean while the large unhatted error bars show the 10 to 90 percentile range of the measured values.}
\label{fig:int_mu}
\end{figure}

Due to equator-ward migration of the appearance of sunspots during a solar cycle, any center-to-limb variation of umbral intensity must be addressed. \inlinecite{albr84} studied the umbra/photosphere intensity ratio in the spectral region 0.387\,--\,2.35 $\mu$m for ten large sunspots (with umbra semi-major axis $\geq10''$) as they rotated from the solar limb to disk center.  They observed a decrease in umbral intensity in the nine (of ten) longer wavelengths, which supports the existence of umbral limb-darkening that theoretical models, such as \inlinecite{avrett81}, predict.  As expected, the darkening is more substantial at longer wavelengths.

In contrast, \inlinecite{nort04} observed two different umbrae with MDI as they traversed across the disk and found an unpredictable yet general \textit{decrease} in umbral intensity towards disk center. As a result, they chose not to employ a correction for center-to-limb variations and point out that sunspot evolution contributes significantly to changes in umbral intensity.  It is worth noting, however, that \inlinecite{albr84} report a very small center-to-limb dependence of umbral intensity at a wavelength comparable to that observed by MDI.  \inlinecite{mathew07} take the more conservative approach of limiting their study of sunspot umbrae to only those near disk center ($\mu\geq0.94$, i.e. latitudes $\leq20$ degrees).  This stringent limit, however, improperly characterizes solar activity over the solar cycle, for sunspots appear as high as 50 degrees latitude ($\mu\approx0.65$), in particular during the beginning of the solar cycle.

Our analysis of umbrae with positions of $\mu\leq0.65$ includes the higher latitude umbrae and thus requires a correction for umbral limb-darkening.  To do this we adopt an approximating premise for the effect of evolution on umbral intensity. We assume that sunspot evolution can be removed through the use of a single relationship between umbral size and minimum umbral intensity.  This single relationship, which is expressed as a quadratic function of umbral radius, is calculated using only umbrae near disk center ($\mu>0.94$) in a manner similar to that discussed in Section~\ref{s:size_int}.  The temporal stability of this relationship is reassuring to our premise.  We illustrate umbral limb-darkening by taking a ratio of the expected umbral intensity, which is calculated singly from the measured umbral size, to the measured umbral intensity.  This is shown in Figure~\ref{fig:int_mu} for all 12967 umbrae.  A linear least-squares fit to the binned ratios can be written as
\begin{equation}
{\frac{I_{\rm calc}}{I_{\rm meas}}= (0.885\pm0.004) + (0.127\pm0.005)\mu}\label{eq:int_mu}
\end{equation}
where I$_{\rm calc}$ is the intensity inferred from the measured umbral radius and I$_{\rm meas}$ is the measured intensity.  Also plotted in the figure is the normalized relationship given for 8760 \AA\ by \inlinecite{albr84}, which agrees remarkably well with the SPM data.  This is the first such confirmation of the results published by \citeauthor{albr84} taking into account variations in umbral size.  We use Equation~\ref{eq:int_mu} as the limb-darkening correction curve for all umbrae.

\subsection{Magnetic Field\,--\,Intensity Relationship}\label{s:mag_int}

\begin{figure}
\centerline{\includegraphics[width=0.95\textwidth,clip=]{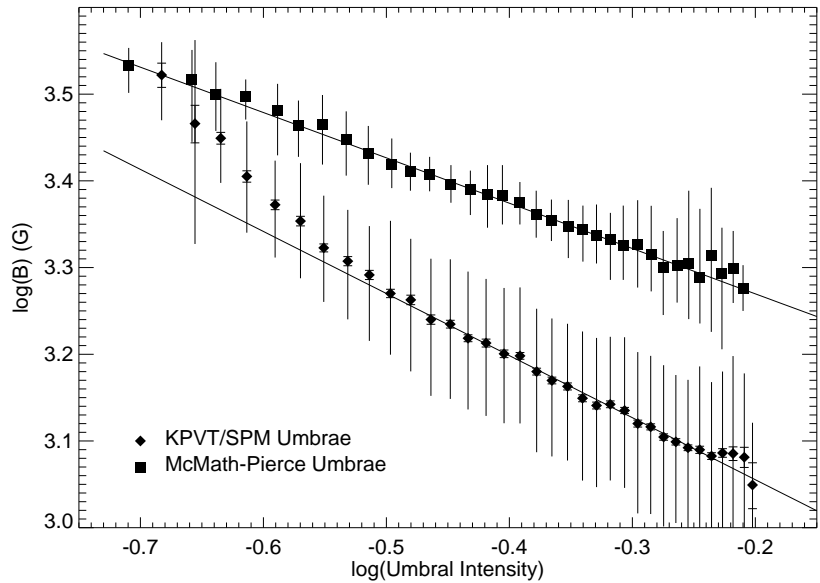}}
\caption{Maximum magnetic field strength versus umbral intensity.  Values for the magnetic field of the selected KPVT/SPM umbrae are position normalized according to Eq.~\ref{eq:vert_b} and shown with diamond data points.  Field values from the McMath-Pierce data (square data points) represent the true magnetic field. The umbral intensity for these values are corrected for the change in observed wavelength using the Planck function.  The error bars are as those in Fig.~\ref{fig:int_mu}.}
\label{fig:mag_int}
\end{figure}

In Figure~\ref{fig:mag_int} we present the observed relationship between the minimum continuum intensity and corresponding magnetic flux for the 9950 observed umbrae that are not close to the solar limb ($\mu>$ 0.65) and have at least a 1.5$''$ radius.  The data is binned into bins 0.0125 (units of contrast) wide. We give the log--log relation to illustrate its linearity (with the exception of darker spots), which is consistent with \inlinecite{nort04}. In addition to the SPM values, the plot shows the up-to-date data taken by Bill Livingston at the McMath-Pierce telescope (McM/P) in the 1.5649 $\mu$m Fe {\sc i} absorption line, which boasts a Land\'e g factor of 3. An earlier subset of these measurements were dicussed in \inlinecite{livingston02}.  Before inclusion in Figure~\ref{fig:mag_int}, we correct for the difference in observed wavelength between the two instruments using the Planck function.  We assume a quiet photospheric temperature of 6000 K for the continuum near each wavelength. The temperature of the umbral core for both wavelengths is also assumed to be identical.  We calculate the umbral temperature at 1.5649 $\mu$m using the Planck function and then use this temperature to calculate the normalized intensity at 8688 \AA.  As the McM/P magnetic measurements reflect the true field, we do not normalize the values for position. We find a general correspondance between the two data sets.  However, there is a large observed difference in magnetic field magnitude, which can be explained by a couple of factors. First, the 1.5649 $\mu$m Fe I line used by Livingston is formed deeper in the atmosphere at a height approximately 110 km above the surface \cite{kopp92} whereas the 8688 \AA\ line is formed approximately 510 km above the surface \cite{giovanelli78}. Furthermore, the KPVT/SPM only measures the line-of-sight component of the field which may underestimate the true field strength which is measured with the fully split McM/P line. The largest factor, however, is the liklely underestimation of the field by the KPVT/SPM as discussed in Section~\ref{s:data}. This too probably explains the difference in slope of the log-log SPM relationship, as well as the upward turning trend for darker spots. The scatter of these plots is well discussed by \inlinecite{livingston02}.

None of these factors affect the $B_{\rm max}$-I$_{\rm min}$ relation over time.  We divide the SPM data set into five different time periods to assess the temporal stability of the B$_{\rm max}$-I$_{\rm min}$ relationship.  A least-squares power law fit is obtained for each time period for spots with umbral intensity values greater than 0.3 (log$(0.3)\approx-0.52$).  The results are presented in Table~\ref{tbl:magvsint}.  For each fit the power law exponent is within a 1.5 sigma deviation of the other fits, thus attesting to its steadiness over time.  While the inclusion of the darker spots does affect the simple power law relationship, they do not discount this result.  We similarly analyze the McMath-Pierce data set to verify this result.  Splitting this data set into three time periods yields a temporally stable relationship just as in the KPVT/SPM data set.

\begin{table}
\caption{Magnetic Field vs. Intensity Power Law Fit Parameters}\label{tbl:magvsint}
\begin{tabular}{lcccc}
\hline
Cycle Phase & N spots & Const. & Exponent & $\chi^{2}$/N \\ 
\hline
KPVT/SPM Data &&& \\
\hline
Falling (Mar 93 - Sep 96)  &  884  & 855$\pm$20.1 & -0.680$\pm$0.029 & 0.0319 \\
Rising (Oct 96 - Mar 01)   & 3832  & 824$\pm$10.0 & -0.731$\pm$0.015 & 0.0096 \\
Falling (Apr 01 - Sep 03)  & 3360  & 797$\pm$9.9  & -0.711$\pm$0.015 & 0.0090 \\
Minimum (1995 - 1997)      &  435  & 858$\pm$29.4 & -0.703$\pm$0.042 & 0.0909 \\
Maximum (2000 - 2002)      & 4268  & 813$\pm$8.9  & -0.710$\pm$0.013 & 0.0064 \\
All data (1993 - 2003)     & 8076  & 817$\pm$6.8  & -0.716$\pm$0.010 & 0.0042 \\
\hline
McMath-Pierce IR Data &&& \\
\hline
Falling (Apr 01 - Jun 05)  &  698 & 1465$\pm$22.6 & -0.520$\pm$0.014 & 0.0291 \\
Maximum (2000 - 2002)      &  156 & 1317$\pm$144  & -0.612$\pm$0.083 & 0.0364 \\
Minimum (2005 - 2007)      &  340 & 1467$\pm$77.6 & -0.522$\pm$0.071 & 0.0084 \\
All data (Apr 90 - Mar 08) & 1084 & 1463$\pm$13.3 & -0.523$\pm$0.009 & 0.0270 \\
\hline
\end{tabular}
\end{table}

\subsection{Magnetic Field\,--\,Size Relationship}

\begin{figure} 
\centerline{\includegraphics[width=0.95\textwidth,clip=]{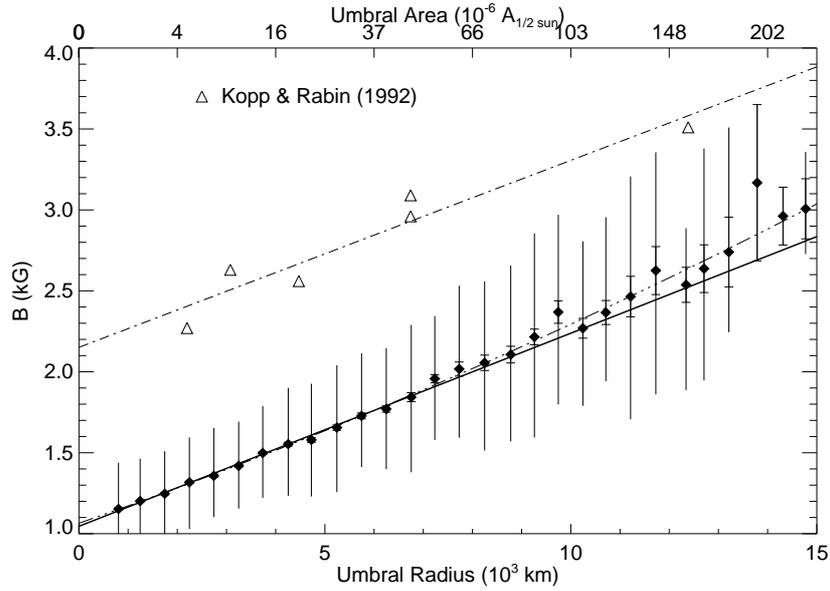}}
\caption{Maximum sunspot magnetic flux versus umbral radius. The diamonds denotes the mean of the binned umbrae from the KPVT/SPM, for which the LOS magnetic flux density values are normalized by position according to Equation~\ref{eq:vert_b}.  Smaller, hatted error bars give the standard error of the mean while the larger error bars illustrate the 10- to 90- percentile range of the points in each bin. The large triangles relay values given by Kopp \& Rabin (1992) for six studied spots. Linear fits are provided as well as a quadratic fit of the SPM data.}
\label{fig:size_mag} 
\end{figure}

In Figure ~\ref{fig:size_mag}, the maximum umbral magnetic field strength is shown as a function of umbral radius for all umbrae with $\mu\geq$ 0.65. Despite the previously discussed error in field magnitude, the relation's positive correlation agrees well with the relation shown for six spots by \inlinecite{kopp92}, which are also shown in the figure. As these IR data measure the true magnetic field strength and not the LOS magnetic flux density, they are not normalized by position.

Table~\ref{tbl:magvsrad} shows the linear least-squares fit parameters for both the SPM data set and the six spots reported by \citeauthor{kopp92}.  Simple linear fits all agree both in time and between the two data sets.  Quadratic fits improve the reduced chi-squared value of the fits for the SPM data and still agree at different phases of the solar cycle.  For a quadratic function of the form
\begin{equation}\label{eq:size_mag}
{B_{r}(R)=B_{0}+bR+cR^{2},}
\end{equation}
the best fit gives the following values: $B_{0}=1064\pm6.7$, b$=105.7\pm3.9$, and c$=1.7\pm0.45$.  The error reflects both the regression error and the statistical error of the magnetic field values.  This curve is also plotted in Figure~\ref{fig:size_mag}.

\begin{table}[t]
\caption{Magnetic Field vs. Radius Linear Fit Parameters}\label{tbl:magvsrad}
\begin{tabular}{lcccccc}
\hline
Cycle Phase & N spots & Constant & Gradient & $\sigma$ & $\chi^{2}$/N \\ 
\hline
Falling (Mar 93 - Sep 96) &  1081 & 1052 & 120.5 & 3.8 & 0.067 \\
Rising (Oct 96 - Mar 01)  &  4876 & 1073 & 118.4 & 1.9 & 0.020 \\
Falling (Apr 01 - Sep 03) &  4369 & 1002 & 123.9 & 1.9 & 0.022 \\
Minimum (1995 - 1997)     &   557 & 1064 & 115.5 & 5.5 & 0.095 \\
Maximum (2000 - 2002)     &  5546 & 1037 & 120.6 & 1.8 & 0.012 \\
All data (1993 - 2003)    & 10326 & 1045 & 119.7 & 1.4 & 0.007 \\
\hline 
Kopp $\&$ Rabin Data      &     6 & 2151 & 115.5 &16.9 & 175.2 \\  
\hline
\end{tabular}
\end{table}

\subsection{Intensity - Size Relationship}\label{s:size_int}

\begin{figure}[t]
\centerline{\includegraphics[width=0.95\textwidth,clip=]{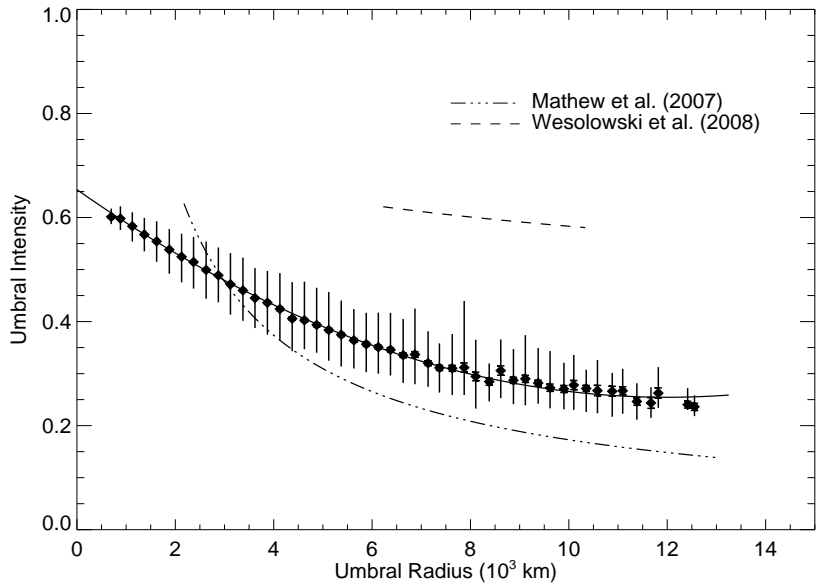}}
\caption{Umbral intensity shown as a quadratic function (solid line) of umbral radius. Each diamond symbol gives the mean umbral intensity for umbrae within a radius bin 250 km wide. The vertical errors bars illustrate the 10- to 90- percentile range of each bin. The standard error of each mean is small. Included are similar functions determined by Mathew \textit{et al.} (2007) and Wesolowski \textit{et al.} (2008) which have been extrapolated to the KPVT/SPM wavelength through the use of the Planck function.  We use the mean solar semidiameter to convert these fits angular measurements to physical units as they make no correction for solar emphemeris.}
\label{fig:size_int}
\end{figure}

The minimum umbral intensity versus umbral radius is presented in Figure~\ref{fig:size_int}.  The data are binned by radius into bins of equal width (250 km) before any regression analysis is attempted.  Thus, each data point represents the mean value of the umbral intensities in a given bin. The small hatted error bars give the standard error of each mean while the larger unhatted error bars show the observed spread of the data by illustrating the 10- to 90- percentile range of the data.  The standard error of each mean is typically small and is propagated into the regression analysis. We find that a 2nd order polynomial, as opposed to a power law relation, lowers the reduced chi-square of the fit.  Similar to the previous sections, this analysis is performed for various portions of the solar cycle as well as the whole data set.  No significant variation of the relationship is found over the solar cycle, which agrees with the analysis of \inlinecite{penn07} and \inlinecite{mathew07}.  For a quadratic fit of the form
\begin{equation}\label{eq:size_int} 
{I_{min}(R)=I_{0}+jR+kR^{2},}
\end{equation}
where R is the umbral radius in seconds of arc and I$_{min}$ is normalized to the surrounding photosphere, we find these best fit parameters for R $>$ 13000 km: $I_{0}=6.454(0.005)\times 10^{-1}$, j$=-6.420(0.03)\times 10^{-2}$, and k$=2.621(0.0418)\times 10^{-3}$. The numeric values given in parentheses reflect the one sigma deviation of the fit parameters.  The relationship is also plotted in Figure ~\ref{fig:size_int}. Additionally, the functions determined by \inlinecite{mathew07} and \inlinecite{wes08} are provided in the figure.  As each of these studies observes the continuum intensity at different wavelengths, the functions given by the authors have been altered using the Planck function to reflect an LTE approximation for the expected intensity of each previous study at the KPVT/SPM wavelength (see Section~\ref{s:mag_int}). We find some consistency between the KPVT/SPM data and the stray-light corrected MDI data analyzed by \inlinecite{mathew07}. For larger umbrae, the KPVT/SPM values are larger than the MDI values suggesting a greater influence of stray light on these umbrae.  For smaller umbrae, KPVT/SPM gives a lower mean minimum intensity than the MDI data.  The differences may be due to stray light effects, or different umbral radius threshold intensities, which change with wavelength as well as different author's determinations.  We do not find a meaningful comparison between the presented data and the function provided by \inlinecite{wes08}.

\subsection{Temporal Evolution of Umbrae}\label{s:cycle}

\begin{figure} 
\centerline{\includegraphics[width=0.95\textwidth,clip=]{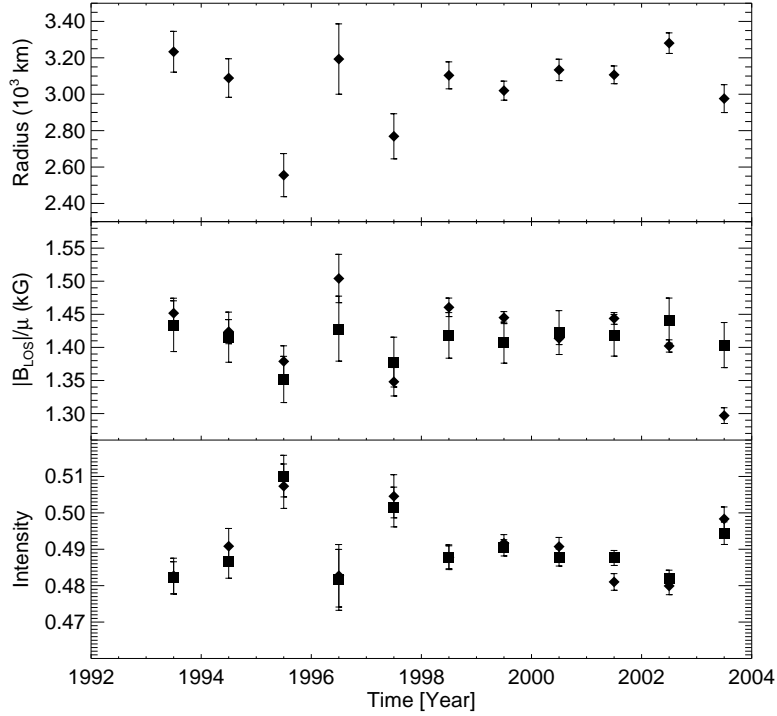}}
\caption{Annual average umbral radius (top), maximum magnetic flux (middle), and intensity (bottom) as a function of time.  Filled diamond data points denote annual averages of the selected umbrae measured by the KPVT/SPM.  Filled square data points illustrate the annual average of the umbral magnetic fluxes and intensities inferred for each individual umbrae from its measured umbral radius and the relationships given in Equations~\ref{eq:size_mag} and~\ref{eq:size_int}.  Error bars represent the standard error of the mean as well as the one sigma error in the determined size relationships.  Mean size variations over the solar cycle correlate with variations in B and I.}
\label{fig:cycle_evol}
\end{figure}

Finally, we address the previous reports of a temporal trend in the mean minimum intensity and mean magnetic field strength by examining these parameters as a function of time in the current data set.  The filled diamonds in Fig.~\ref{fig:cycle_evol} display the mean value of magnetic flux (top) and minimum intensity (bottom) for annual bins between 1993 and 2004.  The error bars give the standard error of each mean.  The oscillation in mean minimum umbral intensity as reported by \inlinecite{penn07} remains evident in our study though the oscillatory amplitude is reduced. A similar oscillation is notable in the mean magnetic field strength in the top plot.  

Although \citeauthor{penn07} found no significant temporal variation in the mean umbral size of their selection, our larger sample size reveals that the selection does exhibit a mean umbral size variation, as shown in Figure~\ref{fig:cycle_evol}.  While the variation is smaller than the error in the size spectrum analysis given in Section~\ref{s:size_dist}, different bins show significant deviation from others.  Putting this to task, we evaluate the expected variation in the umbral minimum intensity and corresponding maximum magnetic flux with time based only on the umbral size and the previously determined time-independent relationships (Equations~\ref{eq:size_mag} and~\ref{eq:size_int}). The correlation coefficient between the measured values of intensity and the calculated ones is 0.91.  Between the observed magnetic flux values and the calculated ones (excluding error), the correlation coefficient is 0.65.  We plot these values in Figure~\ref{fig:cycle_evol} as filled squares.  The error bars associated with these calculated values reflect both the standard error of the annual bin as well as the one sigma error in the determined relationship between umbral radius and the two parameters shown. The one sigma error for the real values and the calculated values are comparable in the figure. For the mean minimum umbral intensity oscillation, we find that changes in umbral radius can account for the perceived oscillation.  And for the magnetic field, all the values calculated from the umbral radii (except 2003) are within the one sigma error of the measured values.  The unchanging relationships of B--R and I--R fully account for the variations of mean umbral intensity and mean magnetic flux over the solar cycle due to a varying mean umbral radius.  Thus, the mean maximum magnetic field strength for the observed umbrae as well as the mean minimum umbral intensity do not give evidence for a solar cycle change in the equilibrium balance between sunspot umbrae and the quiet Sun.

\section{Conclusions}\label{s:conclusions}

Our analysis of more than 10\,000 umbrae observed between 1993 and 2003 gives insight into the temporal stability of bulk umbral structure over the solar cycle, which has been debated since \citeauthor{albr78}'s initial report in 1978 of a cycle dependent change in sunspot intensity.  We find the following:
\renewcommand{\labelenumi}{\textit{\roman{enumi}})} 
\begin{enumerate}
\item The relationships of B--I, B--R, and I--R exhibit power law, linear, and quadratic relationships, respectively, and display no significant modulation during the solar cycle.
\item Minimum intensity measurements support umbral limb-darkening observed in the same spectral region by \inlinecite{albr84} and can be corrected.
\item The umbral size spectrum of the selected umbrae agrees with the conclusion of \inlinecite{bogdan88}.  The umbra size spectrum shape remains constant in time.
\item The periodic variation of the mean umbral intensity and magnetic flux density over the solar cycle can be explained by a corresponding, yet nearly insignificant, change in mean umbral radius.
\end{enumerate} 

Our last result directly conflicts with \inlinecite{penn07} who reported a solar cycle oscillation in mean umbral intensity without a corresponding mean umbral radius oscillation.  That claim relied solely upon no variation observed within annually binned data whose standard errors were large primarily due to a lower number of selected umbrae.  We expand the selection and address the dependencies on umbral radius directly, unlike all reports claiming a solar cycle variation in umbral intensity and/or magnetic field \cite{nort04,penn06,penn07}.  We also test whether our umbral limb-darkening correction and the change in the umbra-penumbra boundary convention affect our result, and no change is evident.

We find the unvarying relationship claimed by \inlinecite{nort04} to be a fairly consistent representation of the bulk magnetohydrostatic equilibrium for sunspots over the solar cycle; however, we also find that considerable scatter plagues the robust use of this relationship even for infrared measurements.  Along with the lack of a significant change in the size distribution of umbrae generated by the Sun, we must conclude that no compelling evidence has yet been found to suggest that the equilibrium balance and/or mean umbral strength is influenced by the progression of the solar cycle.  Though, we add that significant annual variations in the mean umbral radius, as testified by the previously reported oscillations in minimum umbral intensity and maximum magnetic field, do exist.  This claim is further supported by \inlinecite{li05} who studied the ratio of the total daily sunspot area to the number of sunspots between 1874 to 2004.  This ratio exhibited a prominent periodicity consistent and in phase with the solar cycle.  Both of these claims may point to an amplification phase of the toroidal field strength followed by a decay phase.  Still, however, no change in the generated size distribution, as described by \inlinecite{bogdan88}, has been found and the oscillations seen in this paper are not large enough to enact a significant change in the observed size spectrum.  

%
%
\begin{acks}
We express thanks to Bill Livingston for providing his infrared measurements for this work.  Thanks are also extended to Harry Jones with whom we discussed the operation and measurements of the KPVT spectromagnetograph and Jack Harvey for his careful reading of the manuscript. We also thank the anonymous referee for helpful suggestions.  The NSO/Kitt Peak data used here were produced cooperatively by NSF/NOAO, NASA/GSFC and NOAA/SEC.  This work was supported by the National Solar Observatory.
\end{acks}

\end{article} 
\end{document}